\documentclass[proceedings]{JHEP} 

\usepackage{epsfig}                   

\conference{Corfu Summer Institute on Elementary Particle Physics, 1998}

\title{Next-to-Leading Order QCD corrections to the Lifetime Difference  
       of $B_s$ Mesons}

\author{Alexander Lenz    
\thanks{in collaboration with 
    Martin Beneke, Theory Division, CERN, CH-1211 Geneva 23, Switzerland; 
    Gerhard Buchalla, Theory Division, CERN, CH-1211 Geneva 23, Switzerland;
    Christoph Greub, Institut f{\"u}r Theoretische Physik, Universit{\"a}t
    Bern, Sidlerstrasse 5, CH-3012 Berne, Switzerland;
    Ulrich Nierste, Fermilab, Theory Division MS106, IL-60510-500, USA.}
        \\
        Technische Universit{\"a}t M{\"u}nchen\\
        Physik Department\\
        D-85747 Garching\\
        Germany\\
        E-mail: \email{alenz@feynman.t30.physik.tu-muenchen.de}}

\abstract{In this talk we present a calculation of the dacay rate 
          difference in the neutral 
          $B_s-\overline{B}_s$ system, $\Delta \Gamma_{B_s}$,
          in next-to-leading order (NLO) QCD. 
          We find a sizeable decrease compared to leading-order
          (LO) estimates: $ (\Delta \Gamma/\Gamma)_{B_s} = 
          (f_{B_s}/210 MeV)^2 [0.006 B(m_b) + 0.150 B_S(m_b) -
          0.0063]$ in terms of the bag parameters $B$ and $B_S$ in the
          NDR scheme. We put special emphasize on the theoretical and
          physical implications of this quantity.}

\begin{document} 

\section{Non-expert-introduction}
As there were many students in the audience we will start with an
elementary introduction.
Neutral mesons are well known from lectures at the university and were
mentioned here several times e.g. in
\cite{alicorfu, brancocorfu, bertolinicorfu, ruecklcorfu}. 
As in the $K$-system we
have in the $B_s$-system {\it flavour eigenstates} which are defined
by their quark content.
\begin{equation}
| B_s \rangle = ( \bar{b}s ) \; \; ; \; \; | \overline{B}_s \rangle 
= ( b \bar{s} ) \, .
\end{equation}
The {\it mass eigenstates} are linear combinations of the flavour 
eigenstates
\begin{eqnarray}
| B_H \rangle & = & p | B_s \rangle  - q | \overline{B}_s \rangle 
\\
| B_L \rangle & = & p | B_s \rangle  + q | \overline{B}_s \rangle 
\end{eqnarray}
with the normalization condition \mbox{$ |p|^2 + |q|^2 =1$}.
$B_H$ and $B_L$ are the physical states. They have definite masses
and lifetimes, but no definte CP-quantum numbers. The mass eigenstates
are in general mixtures of CP-odd and CP-even eigenstates.

The time evolution of the physical states is described by a simple
Schr{\"o}dinger equation
\begin{equation}
i \partial_t \vec{B} = \hat{H} \vec{B}
\end{equation}
with
\begin{equation}
\vec{B} = 
{ |B_{s} \rangle   \choose | \overline{B}_{s} \rangle }
\; ; \;
\hat{H}
= \left(
\begin{array}{cc}
M-\frac{i}{2} \Gamma & M_{12}-\frac{i}{2} \Gamma_{12} \\
M_{12}^{*} - \frac{i}{2} \Gamma_{12}^{*} & M-\frac{i}{2} \Gamma \\
 \end{array}
\right) \; .
\end{equation}
To find the mass eigenstates and the eigenvalues of the mass operator
and the dacay rate operator we have to diagonalize the hamiltonian. We
get
\begin{eqnarray}
\Delta M_B & = & M_H - M_L = 2 Re (Q)
\\
\Delta \Gamma_B & = & \Gamma_L - \Gamma_H = 4  Im (Q) 
\end{eqnarray}
with
\begin{equation}
Q = \sqrt{(M_{12} - \frac{i}{2} \Gamma_{12})
(M_{12}^* - \frac{i}{2} \Gamma_{12}^*)} \; .
\end{equation}
If we neglect CP violation and expand in $m_b^2/m_t^2$ we can
write with a very good precision 
\begin{eqnarray}
\Delta M_B & = & 2 | M_{12} |
\\
\Delta \Gamma_B & = & - 2 \Gamma_{12} \; .
\end{eqnarray}
The different neutral meson systems gave rise to important
contributions to the field of high energy physics.
In 1964 Christenson, Cronin, Fitch and Turlay \cite{cpex1}    
discovered indirect 
CP-violation\footnote{Indirect or equivalently mixing induced CP
violation  means that
the physical states $K_{S/L}$ are not pure CP-eigenstates. There is a
big contribution of one CP-parity and a tiny of the opposite CP-parity.
If the small contribution dacays, one
speaks of indirect CP violation.}
in the \mbox{$K^0-\overline{K}^0$} -system.
The mass difference in the \mbox{$B_d-\overline{B}_d$} -system was the first 
experimental hint for a very large top quark mass, before the indirect
determination at LEP and before the discovery at Tevatron.
As $m_t$ is by now quite well known, we can extract the 
CKM parameter $|V_{td} V_{tb}|$ from $\Delta M_{B_d}$.
The determination of the CKM parameters is crucial for a test of our
understanding of the standard model and for the search for new physics.
The mass difference in the \mbox{$B_s-\overline{B}_s$} -system is not measured
yet, but we have a lower limit from which we already get an important
bound on the parameters of the CKM matrix. 
\\
The Heavy Quark Expansion (HQE) is the theoretical framework to handle
inclusive $B$-decays. It allows us to
expand the dacay rate in the following way
\begin{equation}
\Gamma = \Gamma_0 
+ \left( \frac{\Lambda}{m_b} \right)^2 \Gamma_2
+ \left( \frac{\Lambda}{m_b} \right)^3 \Gamma_3
+ \cdots \; .
\label{hqe}
\end{equation}
Here we have an systematic expansion in the small parameter
$\Lambda/m_b$. The different terms have the following physical 
interpretations:
\begin{itemize}
\item $\Gamma_0$: The leading term is described by the decay of a free
                  quark (parton model),
                  we have no non-perturbative corrections.

\item $\Gamma_1$: In the derivation of eq. (\ref{hqe}) we make an 
                  operator product expansion. From dimensional reasons
                  we do not get an operator which would contribute to
                  this order in the HQE. \footnote{Strictly spoken we
                  get one operator of the appropriate dimension, 
                  but with the equations of motion we can
                  incorporate it in the leading term.}
      
\item $\Gamma_2$: First non-perturbative corrections arise at the
                  second order in the expansion due to the
                  kinetic and the chromomagnetic operator. They can be
                  regarded as the first terms in a non-relativistic
                  expansion.

\item $\Gamma_3$: In the third order we get the so-called weak
                  annihilation and pauli interference diagrams. 
                  Here the spectator quark is included for the first
                  time .
                  These diagrams give rise to lifetime differences in
                  the neutral $B$-system.
\end{itemize}
Each of these terms can be expanded in a power series in the
strong coupling constant
\begin{equation}
\Gamma_i =  \Gamma_i^{(0)} + 
\frac{\alpha_s}{\pi}   \Gamma_i^{(1)} + \cdots \; .
\end{equation}
So $\Delta \Gamma_B $ has the following form
\begin{equation}
\Delta \Gamma_B = 
\frac{\Lambda^3}{m_b^3}
\left( \Gamma_3^{(0)} + 
\frac{\alpha_s}{\pi}   \Gamma_3^{(1)} + ... \right)
+ \frac{\Lambda^4}{m_b^4}
\left( \Gamma_4^{(0)} +  ... \right) \; .
\end{equation}
After this short introduction for non-experts we motivate the special
interest in the quantity $\Delta \Gamma_{B_s}$.
\section{Motivation}
From a physical point of view one wants to know the exact value of the
decay rate difference, because 
\begin{itemize}
\item $  (\Delta \Gamma / \Gamma)_{B_s}$ is expected to be large. LO
      \cite{dgbslo} estimates give values up to $20\%$. This is on the
      border of the experimental visibility \cite{dgbsex};
\item a big value of $\Delta \Gamma_{B_s}$ would enable us to do novel
      studies of CP-violation without the need of tagging
      \cite{dgbscp}. Tagging is a major expermintal difficulty in B-physics;
\item in the ratio $\Delta \Gamma_{B_s} / \Delta M_{B_s}$ some of the
      non-perturbative parameters cancel \cite{bbd, dgbsnlo}.
      So we can get 
      theoretically clean information on {$ \Delta M_{B_s}$} from a
      measurement of $\Delta \Gamma_{B_s}$;
\item the decay rate difference can be used to search for non
      SM-physics. In \cite{dgbsnp} it was shown that 
      $\Delta \Gamma_{new \; physics} \leq \Delta \Gamma_{SM}$.
\end{itemize}
In order to fullfill this physics program we need a relieable
prediction in the standard model. Therefore we need in addition to
the LO estimate $\Gamma_3^{(0)}$, which was calculated in \cite{dgbslo}
\begin{itemize}
\item the $1/m_b $-corrections $\Gamma_4^{(0)}$. They have been
      calculated by \cite{bbd};
\item the non-perturbative matrix elements for the 
      \mbox{$\Delta B = 2$} 
      operators, which arise in the calculation. Here a relieable
      prediction is still missing;
\item the NLO QCD corrections to the leading term in the $1/m_b$
      expansion, $\Gamma_3^{(1)}$. 
      This was the aim of our work \cite{dgbsnlo}. Besides the better
      accuracy and a reduction of the $\mu$ dependence there is a
      very important point:
      NLO-QCD correction are needed for the proper matching of the
      perturbative calculation to lattice calculations.
\end{itemize}
From a technical point of view this calculation was very interesting
because 
\begin{itemize}
\item 
our result provides the first calculation of perturbative QCD
corrections beyond leading logarithmic order to spectator effects in
the HQE. Soft gluon emmision from the spectator $s$ quark leads to
power-like infrared singularities in individual contributions.
As a conceptual test of the HQE the final result has to be infrared
finite \cite{dgbsir}.
\item 
a crucial point in the derivation of the HQE is the validity of the
operator product expansion. This assumption is known under the name
quark hadron duality and can be tested via a comparison of theory and
experiment. A recent discussion of that subject can be found in     
\cite{duality}. 
\end{itemize}
In the next chapter we will describe the calculation.
\section{Calculation}

The width difference in the \mbox{$B^0-\overline{B}^0$} -system is
defined as
\begin{equation}
\Delta \Gamma = \Gamma_L - \Gamma_H = -2 \Gamma_{21} \; .
\end{equation}
The off-diagonal element of decay-width matrix
can be related to the  so-called transition operator ${\cal T}$ via
\begin{equation}
\Gamma_{21} = \frac{1}{2 M_{B_S}} \langle \bar{B}_S | {\cal T} | B_S 
\rangle
\end{equation}
with
\begin{equation}
{\cal T} = \mbox{Im} \; i \int d^4x \; T \; {\cal H}_{eff}(x) {\cal H}_{eff}(0)
\; .
\label{trans}
\end{equation}
In ${\cal T}$ we have a double insertion of the effective
hamiltonian with the standard form \cite{bbl}
\begin{equation}
{\cal H}_{eff} =  \frac{G_F}{\sqrt{2}} V_{cb}^* V_{cs} \left(
\sum \limits_{r=1}^6 C_r Q_r + C_8 Q_8 \right) \; .
\end{equation}
$G_F$ denotes the Fermi constant, $V_{pq}$ are the CKM matrix
elements and $Q_i$ are local $\Delta B = 1$ operators. 
The Wilson coefficients $C_i$ describe the short distance
physics and are known to NLO QCD.
\\
Formally we proceed now with an operator product
expansion of that product of two hamiltonians.
In real life
one has to calculate diagrams of the following form:
\\
\EPSFIGURE[h]{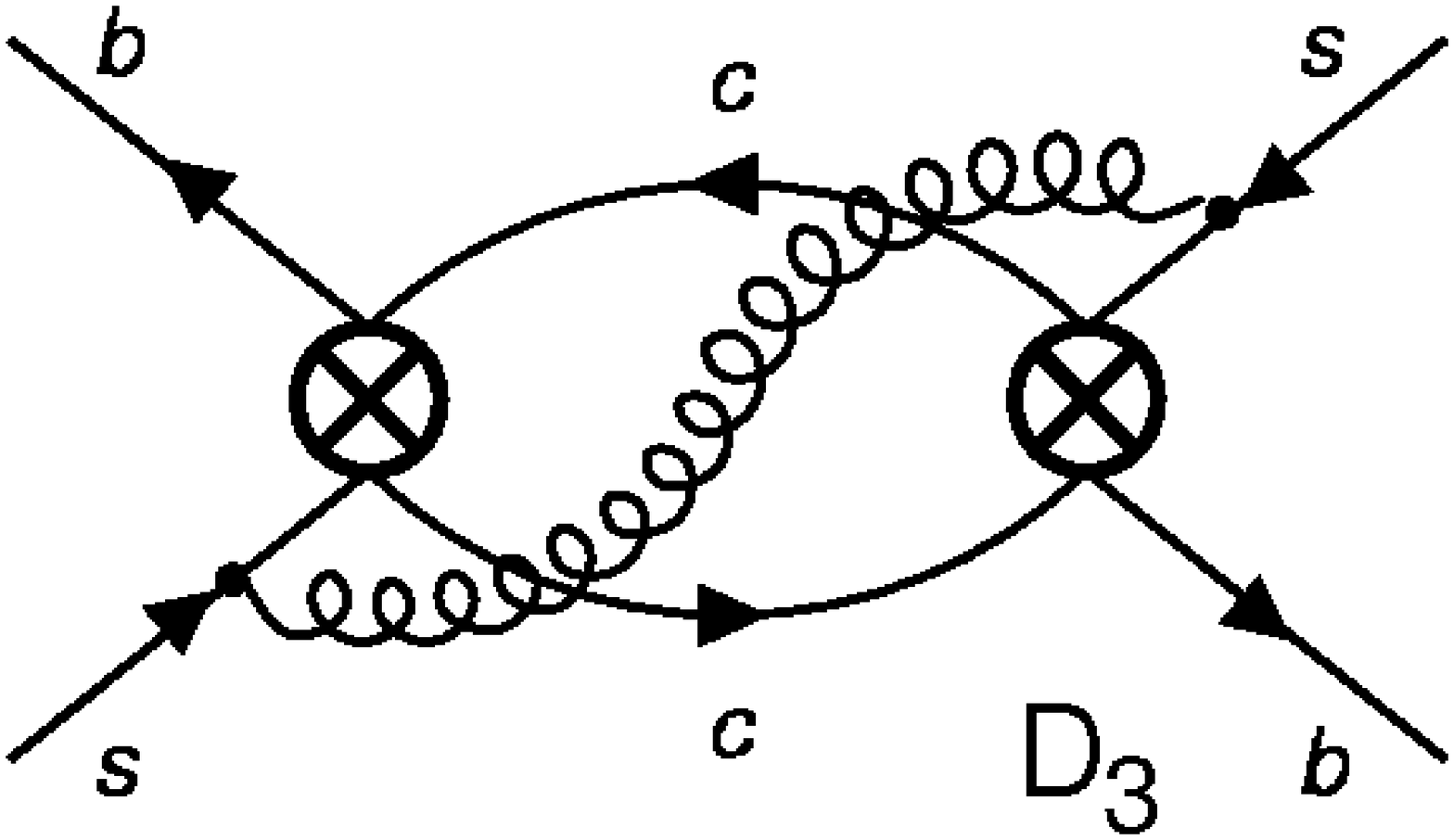, width=5cm}{The imaginary part of massive
two loop diagram of that form has to be calculated.}
\\
One can do the calculation in two different ways (we did it in both
ways, to have a check):
\begin{itemize}
\item calculate  the imaginary part of the two loop integrals
\end{itemize}
or
\begin{itemize}
\item use Cutkosky rules and calculate virtual and real 
one loop corrections, followed by a phase space integration. 
\end{itemize}
The result in LO QCD has the following form
\begin{equation}
{\cal T} = - \frac{G_F^2 m_b^2}{12 \pi} \left( V_{cb}^* V_{cs} \right)^2
\left[ G(z) Q + G_S(z) Q_S \right]
\label{dglo}
\end{equation}
with{  $z = m_c^2/m_b^2$} and the 
$ \Delta B = 2$ operators
\begin{eqnarray}
Q & = & ( \bar{b}_i s_i)_{V-A} ( \bar{b}_j s_j)_{V-A}
\nonumber
\\
Q_S & = &( \bar{b}_i s_i)_{S-P} ( \bar{b}_j s_j)_{S-P} \; .
\end{eqnarray}
In principle we have more operators, but
we can reduce them to the two operators above
with the use of Fierz identities 
\footnote{This reduction is relativeley tricky. For details see
\cite{dgbsnlo}.}.
\\
Equation (\ref{dglo}) is an example of an operator product expansion
of equation (\ref{trans}). We have reduced the double insertion of
\mbox{$\Delta B = 1$} operators, which appear in ${\cal H}_{eff}$,
to a single insertion of an 
\mbox{$\Delta B = 2$} operator. In principle we have integrated out
the internal charm quarks in figure 1.
\\
For the NLO calculation we have to
match the {$ \Delta B = 1$} double insertion with gluon exchange to
a {$ \Delta B = 2$} insertion with gluon exchange. This means, we have
to calculate the following diagrams:
\\
\EPSFIGURE[h]{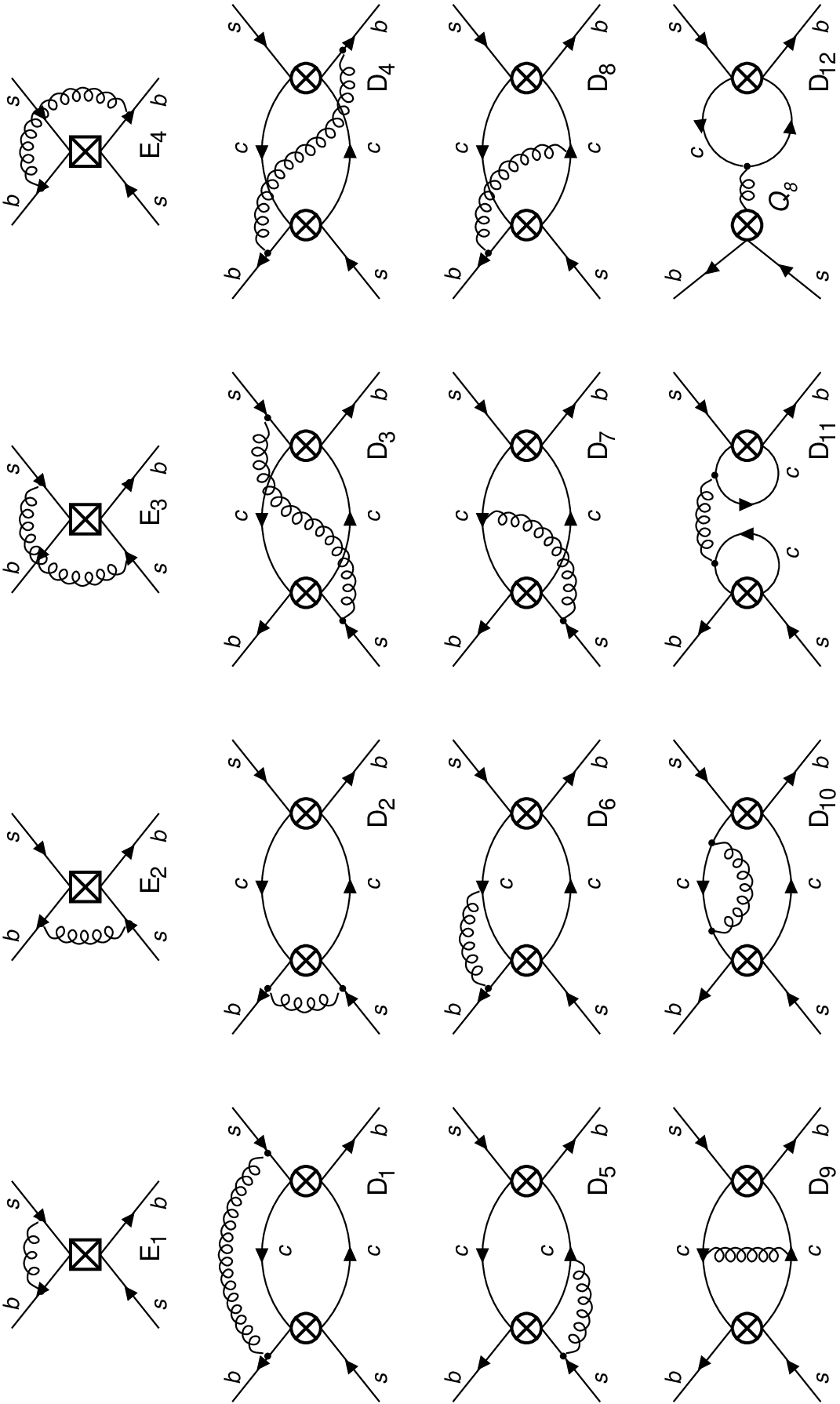, width=4.25cm,angle=270}{All diagrams that
have to be calculated for the NLO QCD determination of $\Delta
\Gamma_{B_s}$.}

These diagrams can be classified in the following way:
\begin{enumerate}
\item[$ E_1 - E_4$:]       Virtual one loop corrections to a 
                           \mbox{$ \Delta B =2 $} operator insertion. 
\item[$ D_1 - D_{10}$:]    Imaginary part of virtual two loop
                           corrections to a double insertion of
                           \mbox{$\Delta B =1 $} operators. 
\item[$ D_{11}, D_{12}$:]  Penguin contributions to the
                           \mbox{$\Delta B =1 $} double insertion.
\end{enumerate}
The calculation of all these diagrams gives us the NLO QCD result.
\section{Results}
The result in NLO is:
{
\begin{equation}
{\cal T} = - \frac{G_F^2 m_b^2}{12 \pi} \left( V_{cb}^* V_{cs} \right)^2
\left[ G(z) Q - G_S(z) Q_S \right]
\end{equation}}
with the following numerical values for the Wilson coefficients 
\begin{displaymath}
\begin{array}{|c|c|c|c|}
\hline
\mu & m_b/2 & m_b & 2 m_b
\\
\hline
G^{(0)} & 0.013 & 0.047 & 0.097
\\
\hline
G & 0.023 & 0.030 & 0.036
\\
\hline
G^{(0)}_S & 1.622 & 1.440 & 1.292
\\
\hline
G_S & 0.743 & 0.937 & 1.018
\\
\hline
\end{array}
\end{displaymath}
with
\begin{equation}
G = G^{(0)} + \frac{\alpha}{4 \pi} G^{(1)} \; .
\end{equation}
Here one can see two important points. First, the value for $G_S$ is
numerical dominant and second, the NLO values are considerably smaller
than the LO values.

For the final result we parametrise the matrix elements of the
\mbox{$\Delta B = 2$} operators in the following way:
\begin{eqnarray}
\langle \bar{B}_s | Q | B_s \rangle & = &
\frac{8}{3} f^2_{B_s} M^2_{B_s} B
\nonumber
\\
\langle \bar{B}_s | Q_S | B_s \rangle & = &
- \frac{5}{3} f^2_{B_s} M^2_{B_s} 
\frac{M^2_{B_s}}{(\bar{m}_b + \bar{m}_s)^2}
B_S \; .
\nonumber
\end{eqnarray}
$B$ and $B_S$ are so-called bag parameters, $f_{B_s}$ is the decay
constant. The values of these parameters have to be determined by 
non-perturbative methods like lattice simulations.
$\bar{m}_q$ denotes the running quark mass in the $\overline{MS}$-scheme.

With the following input parameters
\begin{displaymath}
m_b = 4.8 \mbox{GeV} \; \; \; \; \left( \frac{m_c}{m_b} \right)^2 = 0.085 
\; \; \; \; \bar{m}_s = 0.2 \mbox{GeV}
\end{displaymath}
\begin{equation}
M_{B_s} = 5.37 \mbox{GeV} \; \; \; \;
B ( B_s \mapsto X e \nu ) = 0.104 
\end{equation}
we obtain for the relative dacay rate difference
\begin{equation}
\fbox{$
\begin{array}{ccl}
\left( \frac{ \Delta \Gamma}{\Gamma} \right)_{B_{s}} & = &
\left( \frac{f_{B_s}}{210 \mbox{MeV}} \right)^2
\nonumber
\\
&&
\nonumber
\\
&&
\left[ 0.006 B(m_b) + 0.150 B_S (m_b) - 0.063 \right]
\end{array}
$} \; .
\end{equation}
A definitive determination of the two bag parameters is still missing.
From the literature \cite{lattice} we were able to extract
{\bf  preliminary} values for the bag parameters
\begin{equation}
B (m_b) = 0.9 \; \; \; \; \; \; \; \; B_S (m_b) = 0.75 \; .
\end{equation}
With that numbers at hand
we obtain as a final result
\begin{equation}
\fbox{$ \displaystyle
\left( \frac{ \Delta \Gamma}{\Gamma} \right)_{B_{s}} = 
\left( \frac{f_{B_s}}{210 \mbox{MeV}} \right)^2
\left( 0.054 ^{+0.016}_{-0.032}  \pm ???  \right)
$} \; .
\end{equation}
The question marks remind us that we do not know the uncertainties in
the numerical values for the bag parameters.

\section{Disscussion and outlook}
The LO estimate for the relative decay rate difference 
${ \Delta \Gamma_{B_s} / \Gamma_{B_s} = {\cal{O}} (20\%) }$  is 
considerably reduced due to several effects:
\begin{itemize}
\item the {$ 1/m_b$} corrections are sizeable and give an absolute
      reduction of about { - 6.3 \%} \cite{bbd}.
\item the pure NLO QCD corrections are sizeable, too and give an absolute
      reduction of about { - 4.8 \%} \cite{dgbsnlo}.
\item with the NLO QCD corrections at hand we can perform a proper
      matching to the (preliminary) lattice calculations for the bag
      parameters. This tells us that we have to use a low
      value for the bag parameters, i.e. 
      {$  B_S(m_b) = 0.75$} \cite{dgbsnlo, lattice}. 
      Compared to the naive estimate
      $B_S=1$, this is another absolute reduction of about { - 3.8 \%}. 
\end{itemize}
Unfortunateley the value of $\Delta \Gamma_{B_s}/ \Gamma_{B_s}$ 
has been pinned
down to a value of about $5 \%$. The LO prediction was just a the
border of  experimental visibility \cite{dgbsex}.
Now we will have to wait for the forthcoming experiments like HERA-B,
Tevatron (run II) and LHC.

Another application of our calculation are inclusive indirect
CP-asymmetries in the $ b \to u \bar{u} d$ channel.
For the complete NLO prediction of this quantity, $\Gamma_{12}$ in the
$B_d$ system was missing. We get this value from our calculation with
a trivial exchange of the CKM parameters and the limit $m_c \mapsto 0$.
This allows a determination of the CKM-angle $\alpha$ \cite{dgbsnlo, alpha}.

{\bf Acknowledgements.} I want to thank the organizers of the Corfu
Summer Institute on Elementary Particle Physics for their successful
work, M. Beneke, G. Buchalla, C. Greub and U. Nierste for the pleasant
collaboration and A.J. Buras for proofreading the manuscript.


\begin{thebibliography}{99}
 \bibitem{alicorfu}       A.Ali, these proceedings.
 \bibitem{brancocorfu}    Branco, these proceedings.
 \bibitem{bertolinicorfu} Bertolini, these proceedings.
 \bibitem{ruecklcorfu}    R. R{\"u}ckl, these proceedings.
 \bibitem{cpex1}          J.H. Christenson, J.W. Cronin, V. L. Fitch and
                          R. Turlay, Phys. Rev. Lett. {\bf 13}, (1964), 138.
 \bibitem{dgbslo}         J.S. Hagelin, Nucl. Phys. {\bf B193}, (1981), 123;
			  E. Franco, M. Lusignoli and A. Pugliese, 
			  Nucl. Phys. {\bf B194}, (1982), 403;
			  L.L. Chau, Phys. Rep. {\bf 95}, (1983), 1;
			  A.J. Buras, W. S\l ominski and H. Steger, 
			  Nucl. Phys. {\bf B245}, (1984), 369;
			  M.B. Voloshin, N.G. Uraltsev, V.A. Khoze and
			  M.A. Shifman, Sov. J. Nucl. Phys. {\bf 46},
  			  (1987), 112;
			  A. Datta, E.A. Paschos and U. T{\"u}rke, 
			  Phys. Lett. {\bf B196}, (1987), 382;
                          A. Datta, E.A. Paschos and Y.L. Wu,
                          Nucl. Phys. {\bf B311}, (1988), 35.
 \bibitem{dgbsex}         K. Hartkorn and H.-G. Moser, MPI-PhE/98-21,
                          appears in European Physical Journal {\bf C}.
 \bibitem{dgbscp}         I. Dunietz, Phys. Rev. {\bf D52}, (1995), 3048.
  			  R. Fleischer and I. Dunietz, Phys. Lett.
  			  {\bf B387}, (1996), 361; Phys. Rev. 
			  {\bf D55}, (1997), 259.
 \bibitem{bbd}            M. Beneke, G. Buchalla and I. Dunietz, Phys. Rev. 
                          {\bf D54}, (1996), 4419. 
 \bibitem{dgbsnlo}        M. Beneke, G. Buchalla, C. Greub, A. Lenz and
                          U. Nierste, hep-ph/9808385; to appear in
                          Phys. Lett. {\bf B}.
 \bibitem{dgbsnp}         Y. Grossman, Phys. Lett. {\bf B380}, 99 (1996).
 \bibitem{dgbsir}         I. Bigi and N. Uraltsev, Phys. Lett. 
                          {\bf B280}, (1992), 271.
 \bibitem{duality}        I. Bigi, M. Shifman, N. Uraltsev, A. Vainshtein
                          Phys.Rev. {\bf D59}, (1999), 054011.
 \bibitem{bbl}            G. Buchalla, A.J. Buras and M.E. Lautenbacher,
			  Rev. Mod. Phys. {\bf 68}, (1996), 1125.
 \bibitem{lattice}        R. Gupta, T. Bhattacharya and S.R. Sharpe,
			  Phys. Rev. {\bf D55}, (1997), 4036. 
                          \\
			  J.M. Flynn and C.T. Sachrajda, [hep-lat/9710057],
                          in {\it Heavy Flavours (2nd ed.)}, eds. 
                          A.J. Buras and M. Lindner, World Scientific,
                          Singapore.
 \bibitem{alpha}          M. Beneke, G. Buchalla and I. Dunietz, 
                          Phys. Lett. {\bf B393}, (1997), 132.
\end{thebibliography}
\end{document}